\documentclass[aps,prc,superscriptaddress,showpacs,amsfonts,twocolumn,floatfix]{revtex4-1}

\usepackage{color}
\usepackage{amsmath,amssymb}
\usepackage{epsfig}

\begin{document}


\title{Measurement of the helicity asymmetry $E$ in $\omega\to\pi^+\pi^-\pi^0$ photoproduction}

\newcommand{\comment}[1]{{\color{blue} #1}}

\newcommand*{\ANL}{Argonne National Laboratory, Argonne, Illinois 60439}
\newcommand*{\ANLindex}{1}
\affiliation{\ANL}
\newcommand*{\ASU}{Arizona State University, Tempe, Arizona 85287-1504}
\newcommand*{\ASUindex}{2}
\affiliation{\ASU}
\newcommand*{\BONN}{Helmholtz-Institut f\"ur Strahlen- und Kernphysik, Universit\"at Bonn, 53115 Bonn, Germany}
\affiliation{\BONN}
\newcommand*{\CSUDH}{California State University, Dominguez Hills, Carson, CA 90747}
\newcommand*{\CSUDHindex}{3}
\affiliation{\CSUDH}
\newcommand*{\CANISIUS}{Canisius College, Buffalo, NY}
\newcommand*{\CANISIUSindex}{4}
\affiliation{\CANISIUS}
\newcommand*{\CMU}{Carnegie Mellon University, Pittsburgh, Pennsylvania 15213}
\newcommand*{\CMUindex}{5}
\affiliation{\CMU}
\newcommand*{\CUA}{Catholic University of America, Washington, D.C. 20064}
\newcommand*{\CUAindex}{6}
\affiliation{\CUA}
\newcommand*{\SACLAY}{IRFU, CEA, Universit\'e Paris-Saclay, F-91191 Gif-sur-Yvette, France}
\newcommand*{\SACLAYindex}{7}
\affiliation{\SACLAY}
\newcommand*{\CNU}{Christopher Newport University, Newport News, Virginia 23606}
\newcommand*{\CNUindex}{8}
\affiliation{\CNU}
\newcommand*{\UCONN}{University of Connecticut, Storrs, Connecticut 06269}
\newcommand*{\UCONNindex}{9}
\affiliation{\UCONN}
\newcommand*{\FERRARAU}{Universit\`a di Ferrara, 44121 Ferrara, Italy}
\newcommand*{\FERRARAUindex}{10}
\affiliation{\FERRARAU}
\newcommand*{\FIU}{Florida International University, Miami, Florida 33199}
\newcommand*{\FIUindex}{11}
\affiliation{\FIU}
\newcommand*{\FSU}{Florida State University, Tallahassee, Florida 32306}
\newcommand*{\FSUindex}{12}
\affiliation{\FSU}
\newcommand*{\NRC}{NRC ``Kurchatov Institute", PNPI, 188300, Gatchina, Russia}
\affiliation{\NRC}
\newcommand*{\GENOVAU}{Universit\`a di Genova, Dipartimento di Fisica, 16146 Genova, Italy}
\affiliation{\GENOVAU}
\newcommand*{\GWUI}{The George Washington University, Washington, DC 20052}
\newcommand*{\GWUIindex}{13}
\affiliation{\GWUI}
\newcommand*{\ISU}{Idaho State University, Pocatello, Idaho 83209}
\newcommand*{\ISUindex}{14}
\affiliation{\ISU}
\newcommand*{\INFNFE}{INFN, Sezione di Ferrara, 44100 Ferrara, Italy}
\newcommand*{\INFNFEindex}{15}
\affiliation{\INFNFE}
\newcommand*{\INFNFR}{INFN, Laboratori Nazionali di Frascati, 00044 Frascati, Italy}
\newcommand*{\INFNFRindex}{16}
\affiliation{\INFNFR}
\newcommand*{\INFNGE}{INFN, Sezione di Genova, 16146 Genova, Italy}
\newcommand*{\INFNGEindex}{17}
\affiliation{\INFNGE}
\newcommand*{\INFNRO}{INFN, Sezione di Roma Tor Vergata, 00133 Rome, Italy}
\newcommand*{\INFNROindex}{18}
\affiliation{\INFNRO}
\newcommand*{\INFNTUR}{INFN, Sezione di Torino, 10125 Torino, Italy}
\newcommand*{\INFNTURindex}{19}
\affiliation{\INFNTUR}
\newcommand*{\ORSAY}{Institut de Physique Nucl\'eaire, CNRS/IN2P3 and Universit\'e Paris Sud, Orsay, France}
\newcommand*{\ORSAYindex}{20}
\affiliation{\ORSAY}
\newcommand*{\ITEP}{Institute of Theoretical and Experimental Physics, Moscow, 117259, Russia}
\newcommand*{\ITEPindex}{21}
\affiliation{\ITEP}
\newcommand*{\JMU}{James Madison University, Harrisonburg, Virginia 22807}
\newcommand*{\JMUindex}{22}
\affiliation{\JMU}
\newcommand*{\JINR}{Joint Institute for Nuclear Research, 141980 Dubna, Russia}
\affiliation{\JINR}
\newcommand*{\KNU}{Kyungpook National University, Daegu 41566, Republic of Korea}
\newcommand*{\KNUindex}{23}
\affiliation{\KNU}
\newcommand*{\MISS}{Mississippi State University, Mississippi State, MS 39762-5167}
\newcommand*{\MISSindex}{24}
\affiliation{\MISS}
\newcommand*{\UNH}{University of New Hampshire, Durham, New Hampshire 03824-3568}
\newcommand*{\UNHindex}{25}
\affiliation{\UNH}
\newcommand*{\NSU}{Norfolk State University, Norfolk, Virginia 23504}
\newcommand*{\NSUindex}{26}
\affiliation{\NSU}
\newcommand*{\OHIOU}{Ohio University, Athens, Ohio  45701}
\newcommand*{\OHIOUindex}{27}
\affiliation{\OHIOU}
\newcommand*{\ODU}{Old Dominion University, Norfolk, Virginia 23529}
\newcommand*{\ODUindex}{28}
\affiliation{\ODU}
\newcommand*{\ROMAII}{Universit\`a di Roma Tor Vergata, 00133 Rome Italy}
\newcommand*{\ROMAIIindex}{29}
\affiliation{\ROMAII}
\newcommand*{\MSU}{Skobeltsyn Institute of Nuclear Physics, Lomonosov Moscow State University, 119234 Moscow, Russia}
\newcommand*{\MSUindex}{30}
\affiliation{\MSU}
\newcommand*{\SCAROLINA}{University of South Carolina, Columbia, South Carolina 29208}
\newcommand*{\SCAROLINAindex}{31}
\affiliation{\SCAROLINA}
\newcommand*{\TEMPLE}{Temple University,  Philadelphia, PA 19122 }
\newcommand*{\TEMPLEindex}{32}
\affiliation{\TEMPLE}
\newcommand*{\JLAB}{Thomas Jefferson National Accelerator Facility, Newport News, Virginia 23606}
\newcommand*{\JLABindex}{33}
\affiliation{\JLAB}
\newcommand*{\UTFSM}{Universidad T\'{e}cnica Federico Santa Mar\'{i}a, Casilla 110-V Valpara\'{i}so, Chile}
\newcommand*{\UTFSMindex}{34}
\affiliation{\UTFSM}
\newcommand*{\EDINBURGH}{Edinburgh University, Edinburgh EH9 3JZ, United Kingdom}
\newcommand*{\EDINBURGHindex}{35}
\affiliation{\EDINBURGH}
\newcommand*{\GLASGOW}{University of Glasgow, Glasgow G12 8QQ, United Kingdom}
\newcommand*{\GLASGOWindex}{36}
\affiliation{\GLASGOW}
\newcommand*{\VT}{Virginia Tech, Blacksburg, Virginia   24061-0435}
\newcommand*{\VTindex}{37}
\affiliation{\VT}
\newcommand*{\VIRGINIA}{University of Virginia, Charlottesville, Virginia 22901}
\newcommand*{\VIRGINIAindex}{38}
\affiliation{\VIRGINIA}
\newcommand*{\WM}{College of William and Mary, Williamsburg, Virginia 23187-8795}
\newcommand*{\WMindex}{39}
\affiliation{\WM}
\newcommand*{\YEREVAN}{Yerevan Physics Institute, 375036 Yerevan, Armenia}
\newcommand*{\YEREVANindex}{40}
\affiliation{\YEREVAN}
 
\newcommand*{\NOWUK}{University of Kentucky, Lexington, KY 40506}
\newcommand*{\NOWISU}{Idaho State University, Pocatello, Idaho 83209}

\newcommand*{\UM}{University of Michigan, Ann Arbor, MI 48109}
\newcommand*{\KAERI}{Korea Atomic Energy Research Institute, Gyeongju-si, 38180, South Korea}
\newcommand*{\GABES}{Faculty of Sciences of Gabes, Department of Physics, 6072-Gabes, Tunisia}


\author{Z.~Akbar} \affiliation{\FSU}
\author{P.~Roy} \altaffiliation[Present address: ]{\UM}\affiliation{\FSU}
\author{S.~Park} \altaffiliation[Present address: ]{\KAERI}\affiliation{\FSU}
\author{V.~Crede} \altaffiliation[Corresponding author: crede@fsu.edu]{}\affiliation{\FSU}
\author{A.~V.~Anisovich} \affiliation{\BONN} \affiliation{\NRC}
\author{I.~Denisenko} \affiliation{\BONN} \affiliation{\JINR}
\author{E.~Klempt} \affiliation{\BONN} \affiliation{\JLAB}
\author{V.~A.~Nikonov} \affiliation{\BONN} \affiliation{\NRC}
\author{A.~V.~Sarantsev} \affiliation{\BONN} \affiliation{\NRC}

\author {K.~P.~Adhikari} 
\affiliation{\MISS}
% ...


\collaboration{The CLAS Collaboration}
\noaffiliation

\date{Received: \today / Revised version:}

\begin{abstract}
The double-polarization observable~$E$ was studied for the reaction $\gamma p\to p\,\omega$ 
using the CEBAF Large Acceptance Spectrometer (CLAS) in Hall B at the Thomas Jefferson National 
Accelerator Facility and the longitudinally-polarized frozen-spin target (FROST). The observable was 
measured from the charged decay mode of the meson, $\omega\to\pi^+\pi^-\pi^0$, using a 
circularly-polarized tagged-photon beam with energies ranging from the $\omega$~threshold at 1.1 
to 2.3~GeV. A partial-wave analysis within the Bonn-Gatchina framework found dominant contributions 
from the $3/2^+$~partial wave near threshold, which is identified with the sub-threshold $N(1720)\,3/2^+$ 
nucleon resonance. To describe the entire data set, which consisted of $\omega$~differential cross 
sections and a large variety of polarization observables, further contributions from other nucleon 
resonances were found to be necessary. With respect to non-resonant mechanisms, $\pi$ exchange 
in the $t$-channel was found to remain small across the analyzed energy range, while pomeron
$t$-channel exchange gradually grew from the reaction threshold to dominate all other contributions 
above $W \approx 2$~GeV.
\end{abstract}


\maketitle

\section{\label{Section:Introduction}Introduction}
The production of light vector mesons ($\rho^0$, $\omega$, $\phi$) in 
electromagnetically-induced reactions off the nucleon has attracted interest recently due to the 
availability of high-quality data sets from experiments that study baryon resonances, e.g. at 
Jefferson Laboratory (JLab), the Electron Stretcher Accelerator (ELSA), and the Mainz Microtron 
(MAMI). An experimental program that focuses on the photoproduction of vector mesons at 9~GeV is 
planned at JLab using the GlueX detector~\cite{AlGhoul:2017nbp}. The three lowest-mass vector mesons 
have the same $J^{PC} = 1^{--}$~quantum numbers as the photon. For this reason, the photoproduction 
of these mesons at very high energies, $E_\gamma > 20$~GeV, can successfully be described as a 
diffractive process: The photon converts to a vector meson, which then scatters off the proton 
by the exchange of pomerons. These virtual colorless objects carry no charge and share the 
$J^{PC} = 0^{++}$ quantum numbers of the vacuum~\cite{Donnachie:1994zb}.

At medium energies, $4 < E_\gamma < 20$~GeV, pomeron exchange is not sufficient to describe 
the existing data, e.g. from SLAC~\cite{Ballam:1972eq}, Cornell~\cite{Abramson:1976ks}, 
Daresbury~\cite{Barber:1985fr}, and the exchange of additional Regge families is needed, see e.g. 
the discussion in Ref.~\cite{Laget:2001mu}. The comparison between $\rho$~and $\omega$~data presented in 
Ref.~\cite{Sibirtsev:2003qh} indicated that meson-exchange contributions become important in 
$\omega$~photoproduction. Pion exchange is generally expected to dominate over unnatural exchanges, 
whereas the importance of tensor exchange, which is mediated by the $f_2$ and $a_2$~mesons, is 
{\it a priori} unknown. The authors of Ref.~\cite{Sibirtsev:2003qh} found that in order to describe the 
data, a smooth transition was required between the meson-exchange model at lower energies, $E_\gamma < 5$~GeV, 
and Regge theory at high energies, $E_\gamma > 20$~GeV. They suggested that the dominant contributions 
come from $\pi^0$ and $f_2$-meson exchanges.

Close to the $\omega$~photoproduction threshold in the baryon resonance regime, $N^\ast$~states 
strongly contribute to $\omega$~production. The contributions of twelve $N^\ast$~resonances, along 
with their $N^\ast\to p\omega$~branching ratios, have been determined within the Bonn-Gatchina 
(BnGa) coupled-channels partial-wave analysis (PWA) using data from the CBELSA/TAPS experiment
\cite{Denisenko:2016ugz}. The dominant contribution near threshold was found to be the $3/2^+$
partial wave, which was primarily identified with the sub-threshold $N(1720)\,3/2^+$ resonance. The 
dominance of that partial wave near threshold is surprising since such behavior implies that the decay  
into $\omega N$ proceeds via orbital angular momentum $L=1$. The contributions from the $1/2^-$ 
and $3/2^-$ partial waves were notably smaller, in spite of the fact that the $\omega N$~channel could 
couple to these partial waves with $L = 0$. A significant $3/2^+$~amplitude at low energies was also 
observed in a recent CLAS single-channel PWA, but the authors did not claim any specific resonance 
contributions owing to the complex structure of the $3/2^+$~wave~\cite{Williams:2009aa}. Notable 
contributions from the $5/2^+$ partial wave were reported in both analyses~\cite{Denisenko:2016ugz,
Williams:2009aa}. A structure above $W=2$~GeV has been identified with the $N(2000)\,5/2^+$~state. 
An improved quark model approach to $\omega$-meson photoproduction with an effective Lagrangian 
was presented in Ref.~\cite{Zhao:2000tb}, where the two resonances, $N(1720)\,3/2^+$ and 
$N(1680)\,5/2^+$, were observed to dominate over other excited states.

The isoscalar nature of the $\omega$~meson $(I = 0)$ facilitates the search for nucleon resonances. 
The photoproduction of the $\omega$ in $s$-channel processes can only proceed via $N^\ast$~states 
with $I = \frac{1}{2}$; no contributions from $\Delta^\ast$~resonances with $I = \frac{3}{2}$ are allowed. 

In this paper, we report data obtained for the double-polarization observable known as the helicity 
asymmetry~$E$ for the reaction $\gamma p\to p\omega$, where the $\omega$ was identified through
detection of its decay products $\pi^+\pi^-\pi^0$. The data reported here cover an incident photon 
energy range $E_\gamma$ from 1.1 up to 2.3~GeV, and show (almost) the full angular coverage. The 
observable~$E$ was measured using a circularly-polarized photon beam and a longitudinally-polarized 
proton target. The polarized cross section for this configuration is given by
\begin{equation}
\sigma\,=\,\sigma_0\,(\,1\,-\,\Lambda_z\,\delta_\odot\,E\,)\,,
\end{equation}
where $\sigma_0$ is the unpolarized cross section, $\delta_\odot$ denotes the degree of circular photon-beam 
polarization, $\Lambda_z$ is the degree of longitudinal target-proton polarization, and $E$ is defined as:
\begin{equation}
\label{ReactionE}
\Lambda_z\,\delta_\odot\,E\,=\,\frac{\sigma_{1/2}\,-\,\sigma_{3/2}}{\sigma_{1/2}\,+\,\sigma_{3/2}}\,=
\,\frac{\sigma_{1/2}\,-\,\sigma_{3/2}}{\sigma_0}\,,
\end{equation}
where $\sigma_{1/2}$ and $\sigma_{3/2}$ are the helicity-dependent cross sections with photon and nucleon 
spins anti-aligned and aligned, respectively. 

This paper has the following structure. A brief summary of previous measurements in 
$\omega$~photoproduction is presented in Sec.~\ref{Section:PreviousResults}. 
Section~\ref{Section:ExperimentalSetup} gives an introduction to the CLAS-g9a (FROST) experimental 
setup. The data reconstruction and event selection is discussed in Sec.~\ref{Section:Selection}
and the extraction of the polarization observable is described in Section~\ref{Section:Observable}. 
Finally, the experimental results and a discussion of the observed resonance contributions are presented in 
Secs.~\ref{Section:Results} and~\ref{Section:PWA}.


\section{\label{Section:PreviousResults}Previous Measurements}
Cross section data for the reaction $\gamma p\to p\omega$ were obtained and studied at many different 
laboratories over a wide kinematic range~\cite{Barth:2003kv,Ajaka:2006bn,Williams:2009ab,Wilson:2015uoa,
Strakovsky:2014wja}. A review of the main data sets published before 2013 and a corresponding comparison 
of their coverage in energy and solid angle can be found in Ref.~\cite{Crede:2013sze}. The total cross section 
for $\omega$~photoproduction reaches about $8.5~\mu$b and exhibits a pronounced peak structure at about 
$E_\gamma = 1.3$~GeV in addition to a broader peak at about $E_\gamma = 1.9$~GeV~\cite{Wilson:2015uoa},
similar in shape to that seen in $\rho$ and $\phi$~production. The differential cross sections, d$\sigma$/d$t$, 
show an exponential fall-off at small values of the squared recoil momentum, $t$. 

Few measurements exist for polarization observables in $\omega$~photoproduction. The photon-beam 
asymmetry~$\Sigma$ was first measured by the GRAAL Collaboration in 2006 from the decay modes 
$\omega\to \pi^0\gamma$ and $\omega\to\pi^+\pi^-\pi^0$, and was presented in four energy bins that 
cover an energy range from threshold up to a photon energy of 1.5~GeV~\cite{Ajaka:2006bn}. A second 
measurement based on both decay modes was published in 2015 and showed improved angular coverage
\cite{Vegna:2013ccx}. The photon-beam asymmetry~$\Sigma$ was also measured by the CBELSA/TAPS 
Collaboration for a maximum energy of $E_\gamma = 1.5$~GeV~\cite{Klein:2008aa}. We refer to 
Refs.~\cite{Collins:2017vev,Roy:2017qwv} prepared by the CLAS Collaboration for a detailed discussion 
of $\Sigma_\omega$ including new high-statistics data from Jefferson Lab~\cite{Roy:2017qwv}.

The first measurements of $\omega$~double-polarization observables were reported from the CBELSA/TAPS 
Collaboration~\cite{Eberhardt:2015lwa}. The publication provided five data points for the observable~$G$ 
at a single energy interval of $1108 < E_\gamma < 1300$~MeV and the helicity asymmetry, $E$, for a photon 
energy range from close to threshold at $E_\gamma = 1108$~MeV to $E_\gamma = 2300$~MeV. Both 
measurements cover the full solid angle.

Decays of vector mesons give rise to additional spin observables beyond those accessed in pseudoscalar 
meson decays, and vector-meson decays provide the opportunity to access spin-density matrix elements 
(SDMEs). High-statistics results on SDMEs, $\rho^0_{00}$, $\rho^0_{10}$, and $\rho^0_{1-1}$, have already 
been measured at CLAS~\cite{Williams:2009ab} (for $E_\gamma < 3.8$~GeV) and 
CBELSA/TAPS~\cite{Wilson:2015uoa} (for $E_\gamma < 2.5$~GeV) using an unpolarized photon beam, and 
CBELSA/TAPS~\cite{Wilson:2015uoa} also reported the first measurements of polarized SDMEs ($\rho^1_{00}$,
$\rho^1_{11}$,~Re~$\rho^1_{10}$, and Im~$\rho^2_{1-1}$), using a linearly-polarized photon beam for 
$E_\gamma < 1.65$~GeV. 

\section{\label{Section:ExperimentalSetup}Experimental Setup}
The experiment was performed at the Continuous Electron Beam Accelerator Facility (CEBAF) at Jefferson 
Lab using the CEBAF Large Acceptance Spectrometer (CLAS)~\cite{Mecking:2003zu} in Hall~B with a 
circularly-polarized, tagged, bremsstrahlung photon beam whose helicity state was changed pseudo-randomly 
at a rate of 29.560~Hz. The measurements were part of the ``g9a'' running period, which were the first 
measurements using the Jefferson Lab ``frozen spin'' target FROST~\cite{Keith:2012ad} described below. 
A circularly-polarized photon beam results from a polarization transfer when the incident electron beam 
itself is longitudinally polarized. 
The electron beam polarization was determined with the Hall~B M\o ller polarimeter~\cite{Moeller} that measured 
the asymmetry in elastic electron-electron (M\o ller) scattering. The data for the double-polarization observable~$E$ 
were recorded in seven different groups of runs defined by the target-proton polarization and two different accelerator 
energies with electron-beam polarization degrees, $\delta_{\rm e^-}$, of 84.8\,\% and 83.0\,\%, respectively; the 
uncertainty in the degree of electron-beam polarization was about 1.4\,\%~\cite{Mecking:2003zu}.

\begin{figure}[t]
 \includegraphics[width=0.49\textwidth,height=0.32\textheight]{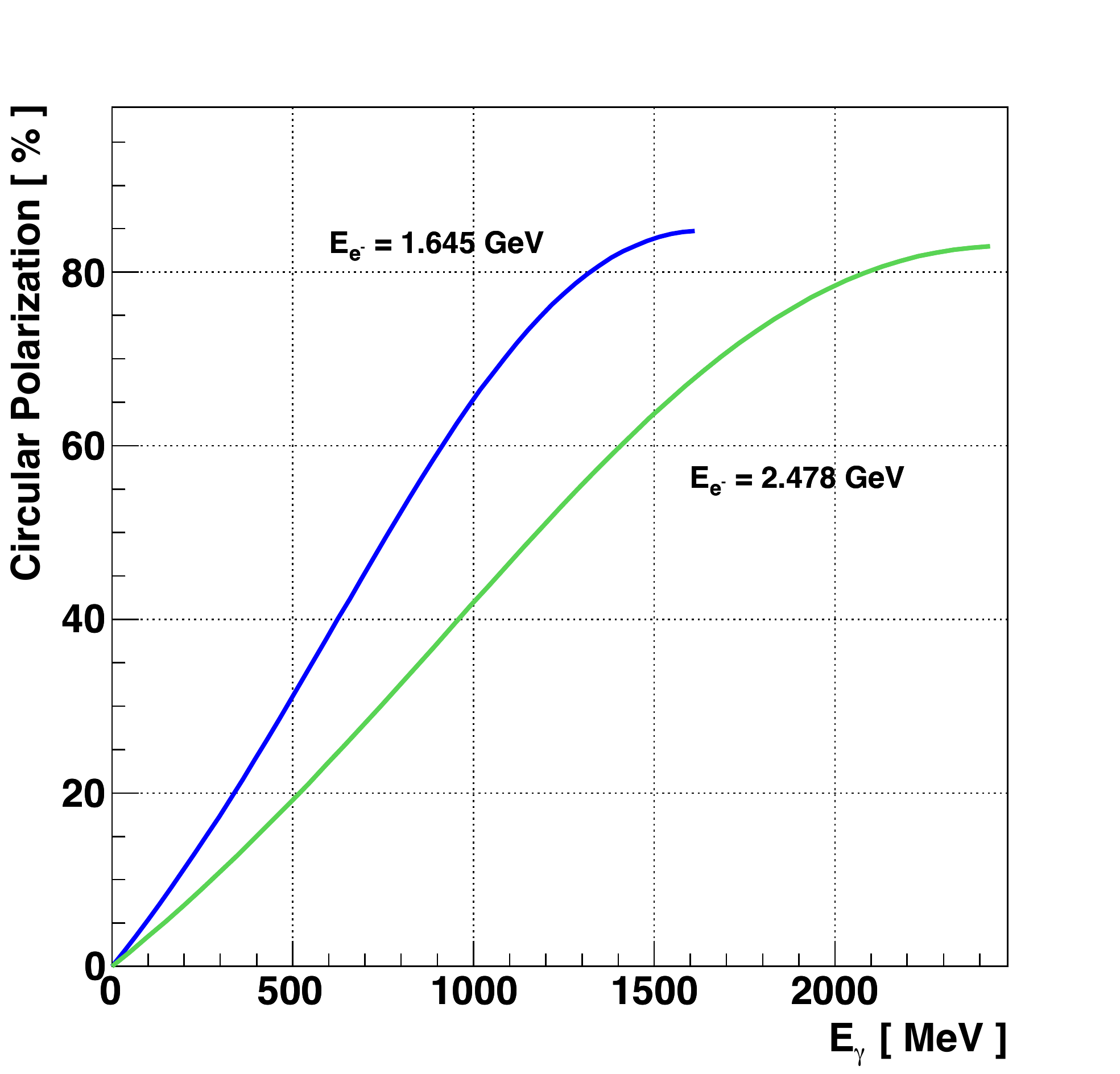}
 \caption{\label{Figure:CircularPolarization}(Color online) Degree of circular-photon polarization as a 
   function of photon energy for the two CEBAF energies of 1.645~GeV (blue) and 2.478~GeV (green).}
\end{figure}

The longitudinally-polarized electron beam was extracted from the CEBAF accelerator and was incident on 
the thin radiator of the Hall B photon tagger~\cite{Sober:2000we}. The photon tagging system included a 
focal plane incorporating a layer of 384~partially overlapping small scintillators that detected electrons 
that had undergone bremsstrahlung; the small scintillators thus provided the photon-beam energy definition 
and resolution via energy conservation. A second layer of 61~larger scintillators provided the timing resolution 
for an event through a coincidence of an electron passing through one of the larger scintillators with the 
detection of decay products following meson photoproduction as described below. 

\begin{figure}[b]
 \includegraphics[width=0.49\textwidth,height=0.33\textheight]{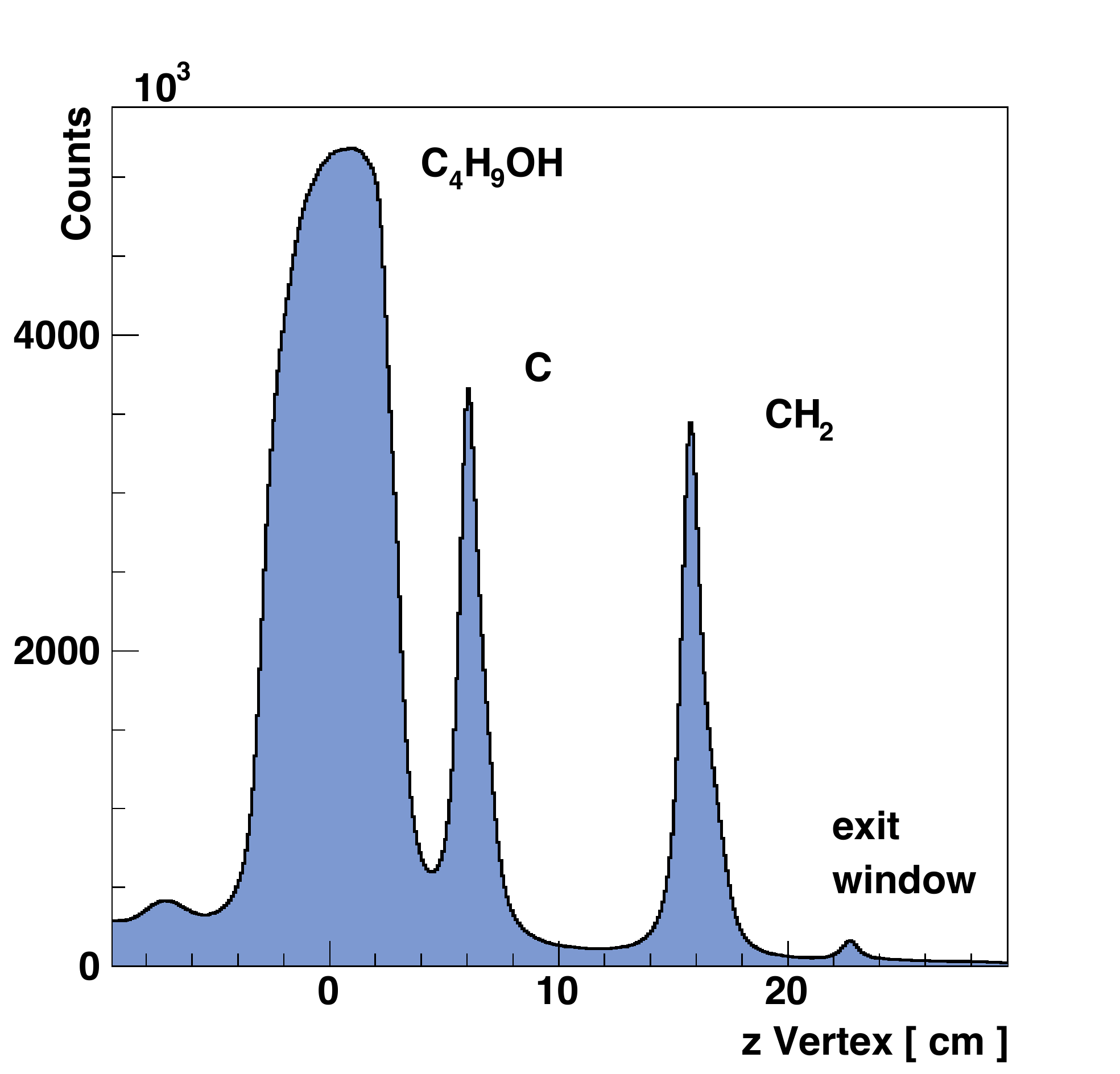}
 \caption{\label{Figure:zVertex}(Color online) The $z$-vertex distribution (axis along the beamline) in
   the FROST-g9a experiment based on about 30\,\% of the total statistics for the full photon energy
   range. The three peaks for the different targets are clearly visible: butanol, carbon, and polyethylene 
   (from left to right). Also visible are the exit window of the vacuum chamber and an enhancement to 
   the left of the butanol peak where the target cup was attached to a stainless steel tube, which was used 
   to insert the cup into the cryostat.}
\end{figure}

In this experiment, the tagging system produced circularly-polarized tagged photons in the energy range 
between $E_\gamma = 0.35$ and 2.37~GeV with an energy resolution of $\sim 10^{-3}\,E_{e^-}$. The degree 
of circular polarization of the bremsstrahlung photons, $\delta_\odot$, was determined from the 
polarization transfer of the longitudinally-polarized electrons~\cite{Olsen:1959zz}:
\begin{equation}
\delta_\odot \,=\, \delta_{\rm e^-} \,\cdot\, \frac{4x\,-\,x^2}{4\,-\,4x\,+\,3x^2}\,,
\end{equation}
where $x = E_\gamma / E_{\rm e^-}$, and $E_\gamma$ as well as $E_{\rm e^-}$ are the energy of the incoming 
photon and the energy of the electron beam, respectively. Figure~\ref{Figure:CircularPolarization} shows that 
the degree of the circular photon-beam polarization is roughly proportional to the photon-beam energy.

\begin{figure*}[t]
 \includegraphics[width=0.32\textwidth]{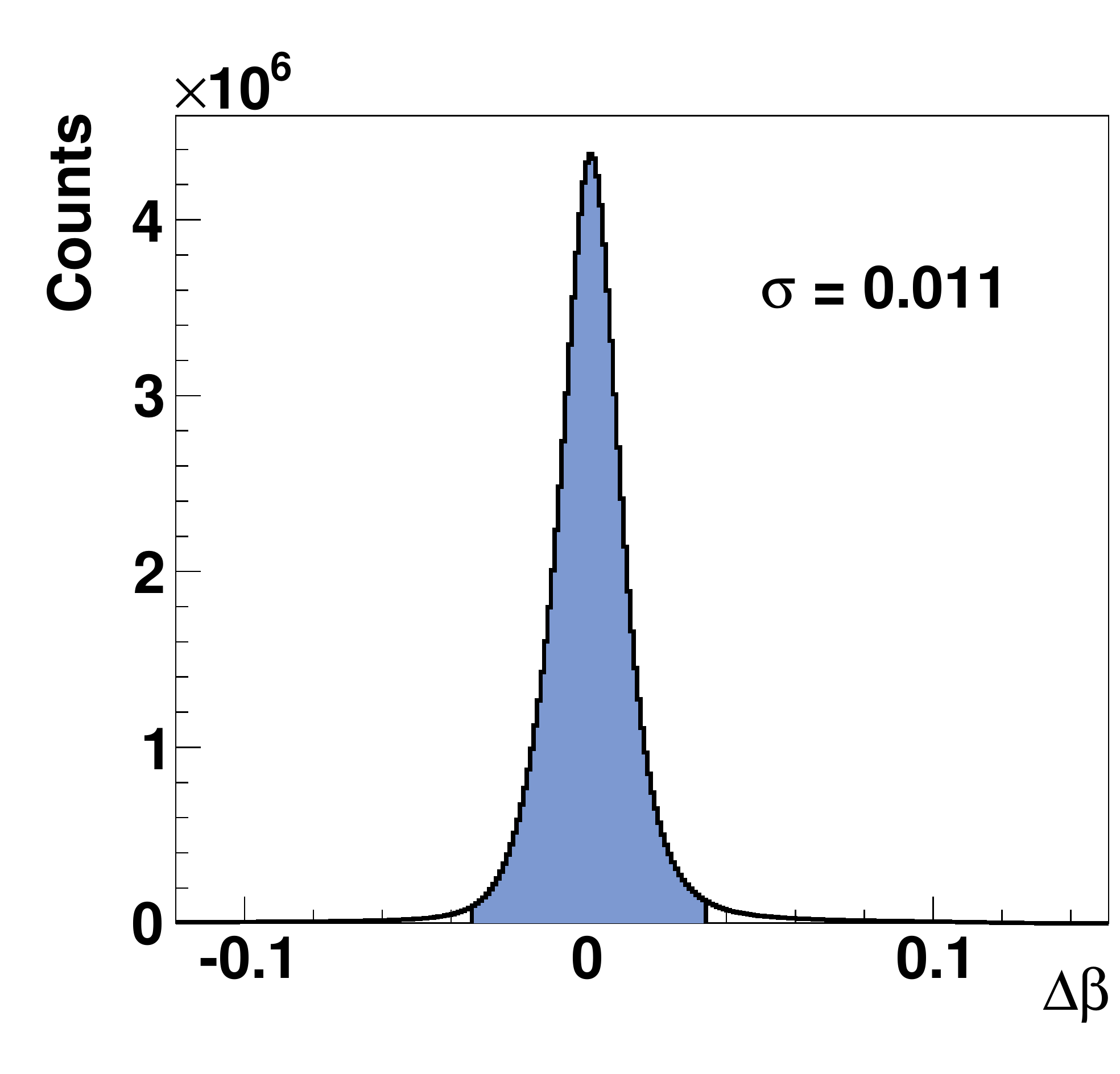}
 \includegraphics[width=0.32\textwidth]{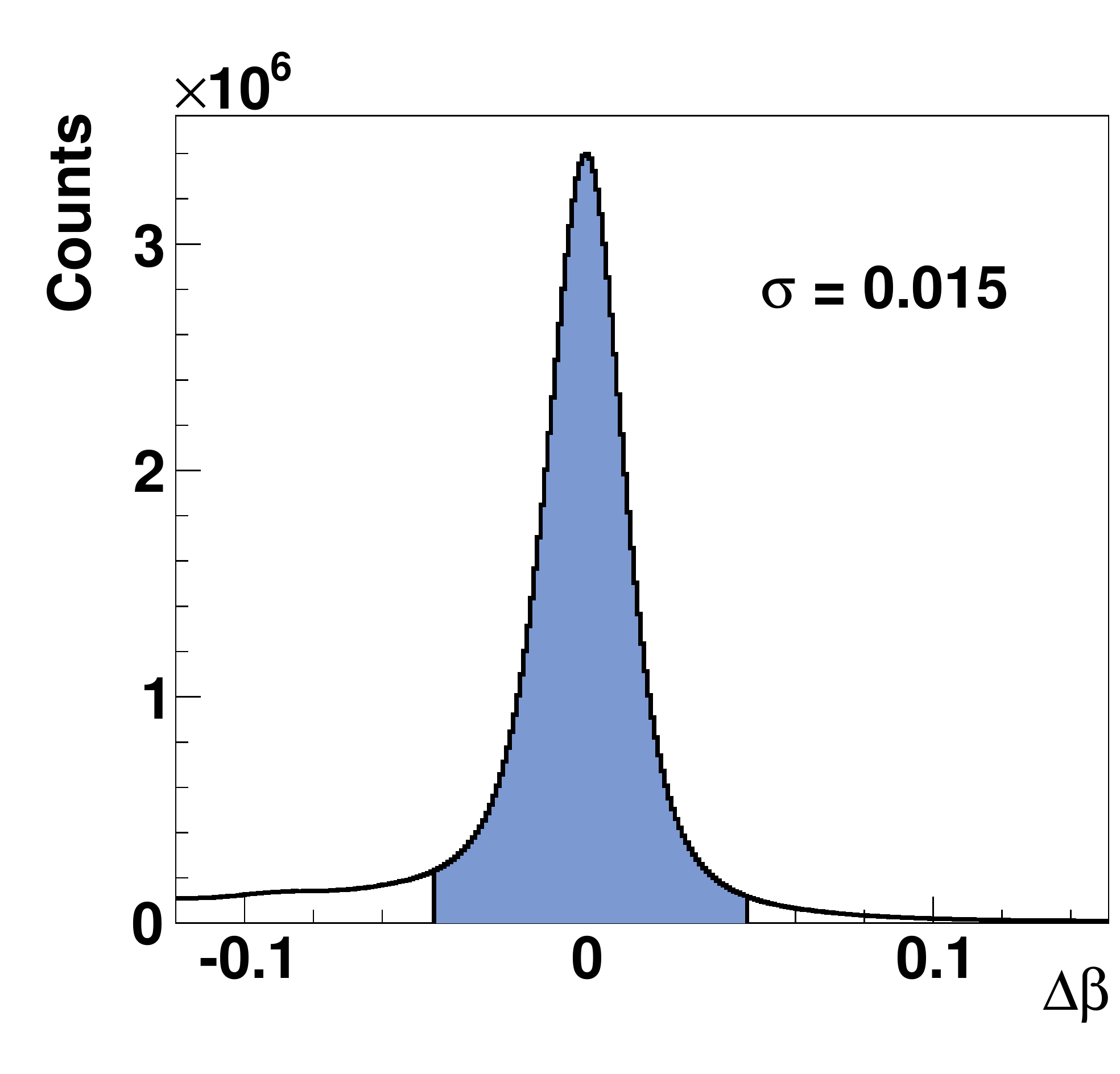}
 \includegraphics[width=0.32\textwidth]{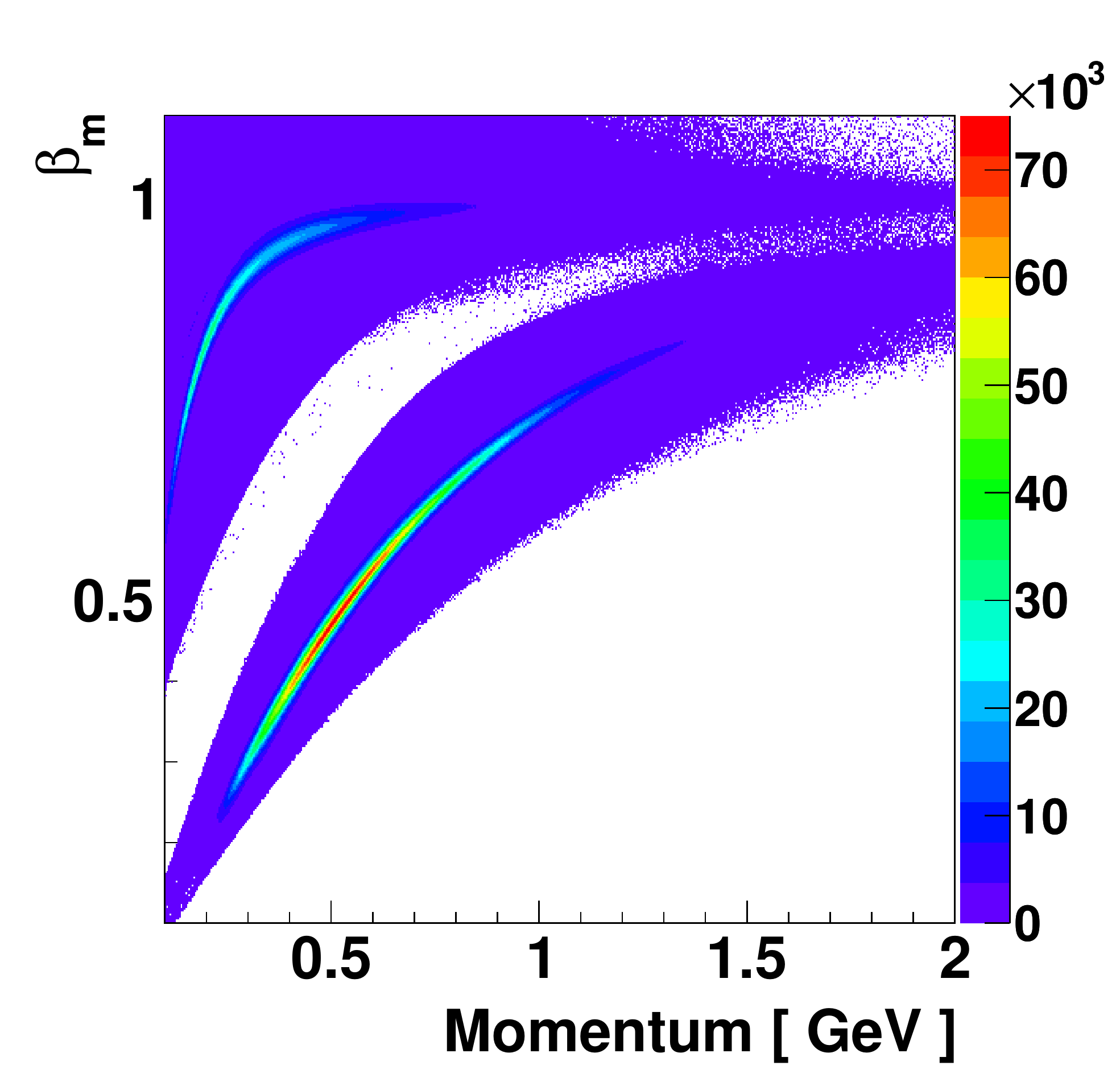}
 \caption{\label{Figure:beta} (Color online) Left and middle: $\Delta\beta\,=\,|\,\beta_c\,-\,\beta_m|$
   distributions for protons and charged pions, respectively. The blue area indicates the $3\sigma$~cuts 
   according to  Eq.~(\ref{Equation:beta}). Right: The distribution of $\beta_m$ versus particle momentum 
   before the $3\sigma$~cuts.}
\end{figure*}

The charged particles in the $p\pi^+\pi^-\pi^0$~final state were detected in the CLAS spectrometer, 
which provided a large coverage for charged particles in the polar-angle range 
$8^\circ < \theta_{\rm lab} < 135^\circ$. The four-momentum vectors of the particles were reconstructed 
from their tracks in the toroidal magnetic field of the spectrometer by a set of three drift-chamber 
packages~\cite{Mestayer:2000we} and by particle identification using time-of-flight information from 
plastic scintillators located about 5~m from the target~\cite{Smith:1999ii}. The CLAS spectrometer 
provided a momentum and angle resolution of $\Delta p/p\approx 1\,\%$ and 
$\Delta\theta\approx 1^\circ$\,-\,$2^\circ$, respectively. A set of plastic scintillation counters 
close to the target provided event start times~\cite{Sharabian:2005kq}. For this experiment, coincident 
signals from the photon tagger, start-, and time-of-flight counters constituted the event trigger that
required a coincidence between a scattered-electron signal from the photon tagger and at least one 
charged track in CLAS. More details on the spectrometer can be found in Ref.~\cite{Mecking:2003zu}.

Data from reactions using the FROzen Spin Target (FROST)~\cite{Keith:2012ad} at the center of the CLAS 
spectro\-meter were accumulated. The target material consisted of frozen beads of butanol (C$_4$H$_9$OH) 
that were 1\,-\,2~mm in diameter. Approximately 5~g of these beads were loaded into a cylindrical target 
cup with a diameter of 15~mm and a length of 50~mm. The target was longitudinally polarized with microwaves 
via Dynamic Nuclear Polarization~(DNP)~\cite{DNP} in the bore of a 5~T polarizing (solenoid) magnet outside
CLAS at about 200\,-\,300~mK. The polarization was maintained at a frozen-spin temperature of about~30~mK 
inside the spectro\-meter by a weaker 0.56~T holding field during data taking. The FROST target was typically 
polarized with an average starting polarization of 84\,\% in the positive-spin state and $-86$\,\% in the negative. 
Relaxation times ranged from about 2800~h with beam on target to about 3600~h without beam. The target relaxed more 
quickly in the negative spin state, about 1400~h with beam and 1900~h without. The maximum polarization was $-94\,\%$. 
The target was re-polarized (and the polarization reversed) about once a week. The design details and the target 
performance in the FROST experiment are discussed in Ref.~\cite{Keith:2012ad}. 

The DNP technique was realized by placing the target material in a high magnetic field under conditions
such that the polarization of the free electron spins approached unity. Spin flips of an electron and that 
of a nearby free proton were induced by microwaves of frequency near the electron spin resonance. Since 
the nucleon spins couple more weakly with the lattice than the electron spins, their spin-relaxation rates 
were much longer and the nucleons could accumulate into either the positive or negative spin state without 
reversing the magnetic field. This could be tuned by the proper microwave frequency. As a result, the 
direction of the target-proton polarization in this experiment was defined by two quantities: The direction 
of the proton polarization with respect to the holding magnetic field and the direction of the holding 
magnetic field with respect to the incident photon-beam polarization plane. The degree of the 
target-proton polarization was measured during the run with the continuous wave nuclear 
magnetic resonance (NMR) technique~\cite{Keith:2012ad}.

Data were also simultaneously obtained from two additional targets: a 1.5-mm-thick carbon disk and a
3.5-mm thick CH$_2$ disk at approximately 6~and 16~cm downstream of the butanol sample, respectively. 
Figure~\ref{Figure:zVertex} shows the $z$-vertex distribution in the FROST-g9a experiment based upon about 
30\,\% of the total statistics. The three dominant peaks for the different targets are clearly visible. The
carbon target was used to study background from bound nucleons and to determine dilution factors, whereas 
the CH$_2$ target provided relevant information on events off unpolarized nucleons. The thickness of the 
additional targets was chosen such that the hadronic rate from each was about 10\,\% the rate of butanol.

\section{\label{Section:Selection}Preparation of Final States}
The data presented here were accumulated between November 2007 and February 2008 in seven run 
periods with CEBAF energies of 1.645~GeV (Periods 1\,-\,3) and 2.478~GeV (Periods 4\,-\,7). These
data were also used to extract the helicity asymmetry for a variety of other final states, see e.g. 
Refs.~\cite{Strauch:2015zob,Senderovich:2015lek}. The event reconstruction and selection of the
photoproduction channel $\gamma p\to p\omega\to p\pi^+\pi^-\pi^0$ is described below and resulted 
in the reconstruction of 62,300 $\omega$ events from the full data set obtained in this experiment.

\subsection{\label{Subsection:Reconstruction}Event reconstruction}
The reaction $\gamma p\to p\pi^+\pi^-\,(\pi^0)$ was identified in a first step by requiring 
exactly one proton track and two charged-pion tracks in the CLAS detector. Positively- and 
negatively-charged pions were distinguished by their track curvatures in the toroidal magnetic 
field. The acceptance of $\pi^-$~mesons was smaller than for $\pi^+$~mesons since they were bent 
toward the beamline and a large fraction escaped through the forward hole of the CLAS spectrometer.

\begin{figure}[t]
 \includegraphics[width=0.47\textwidth,height=0.31\textheight]{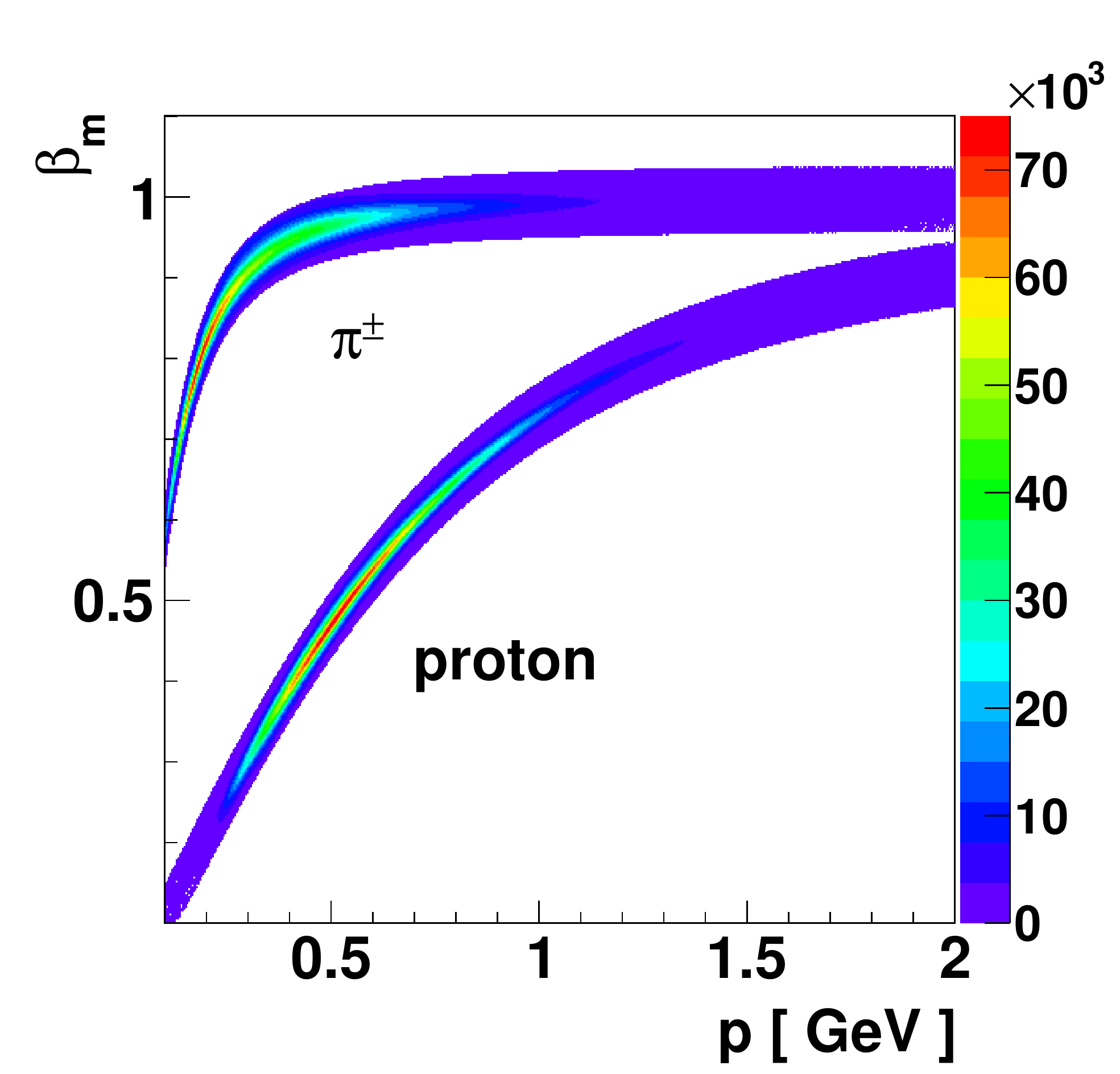}
 \caption{\label{Figure:betaFinal}(Color online) The distribution of $\beta_m$ versus particle momentum 
     after the $3\sigma$~cuts on $\Delta\beta$ according to Eq.~(\ref{Equation:beta}).}
\end{figure}

Particle identification was then improved by applying a cut on $\Delta\beta$:
\begin{equation}
\Delta\beta\,=\,|\,\beta_c\,-\,\beta_m|\,=\,|\,\sqrt{\frac{p^2}{m^2\,+\,p^2}}\,-\,\beta_m\,|\,<\,3\sigma\,,
\label{Equation:beta}
\end{equation}
where $\beta_m = v/c$ was the empirically-measured value for each particle based on timing information from
the time-of-flight and start counter systems, and $\beta_c$ was determined from the measured momentum 
using the CLAS drift chambers and the PDG mass~\cite{Olive:2016xmw} for the particle. While the quantity
$\Delta\beta$ depends on particle momentum, the $\Delta\beta$~distribution is approximately Gaussian when 
summed over all $\beta_m$~values, with width $\sigma = 0.011$ and $0.015$ for the proton and pions, respectively. 
Figure~\ref{Figure:beta} shows the $\Delta\beta$~distributions for protons (left) and charged pions (middle). 
The tail on the left side of the $\Delta\beta$~peak for pions originates from misidentified electrons. Also 
shown in Fig.~\ref{Figure:beta} (right) is the distribution of $\beta_m$ versus particle momentum before the 
$3\sigma$~cuts. The final distribution of $\beta_m$ versus particle momentum after the $3\sigma$~cuts on 
$\Delta\beta$ according to Eq.~(\ref{Equation:beta}) is shown in Fig.~\ref{Figure:betaFinal}. Clear bands 
for the proton and the pions are visible. 

All detected final-state particles were also corrected for their energy loss along the path from the target 
to the time-of-flight scintillator array. Moreover, the magnitude of the particle momentum was corrected for 
small misalignments of the CLAS drift chambers and fluctuations in the toroidal field. These corrections were 
typically of the order of a few MeV.

All detected final-state particles exhibited small modulations in the laboratory polar and azimuthal angular 
distributions with amplitudes of $< 0.5^\circ$. These modulations were consistent with effects of the 
solenoidal holding field on charged particles. The four-momentum for each detected particle  was corrected 
independently in both angles; given the size of the effect, correlations between the two angles were considered 
negligible. Since the $E$~observable was extracted separately from each of the seven groups of runs without mixing 
data using different holding field directions, any remaining effect would drop out when the asymmetries were formed. 

\begin{figure}[b]
 \includegraphics[width=0.46\textwidth,height=0.3\textheight]{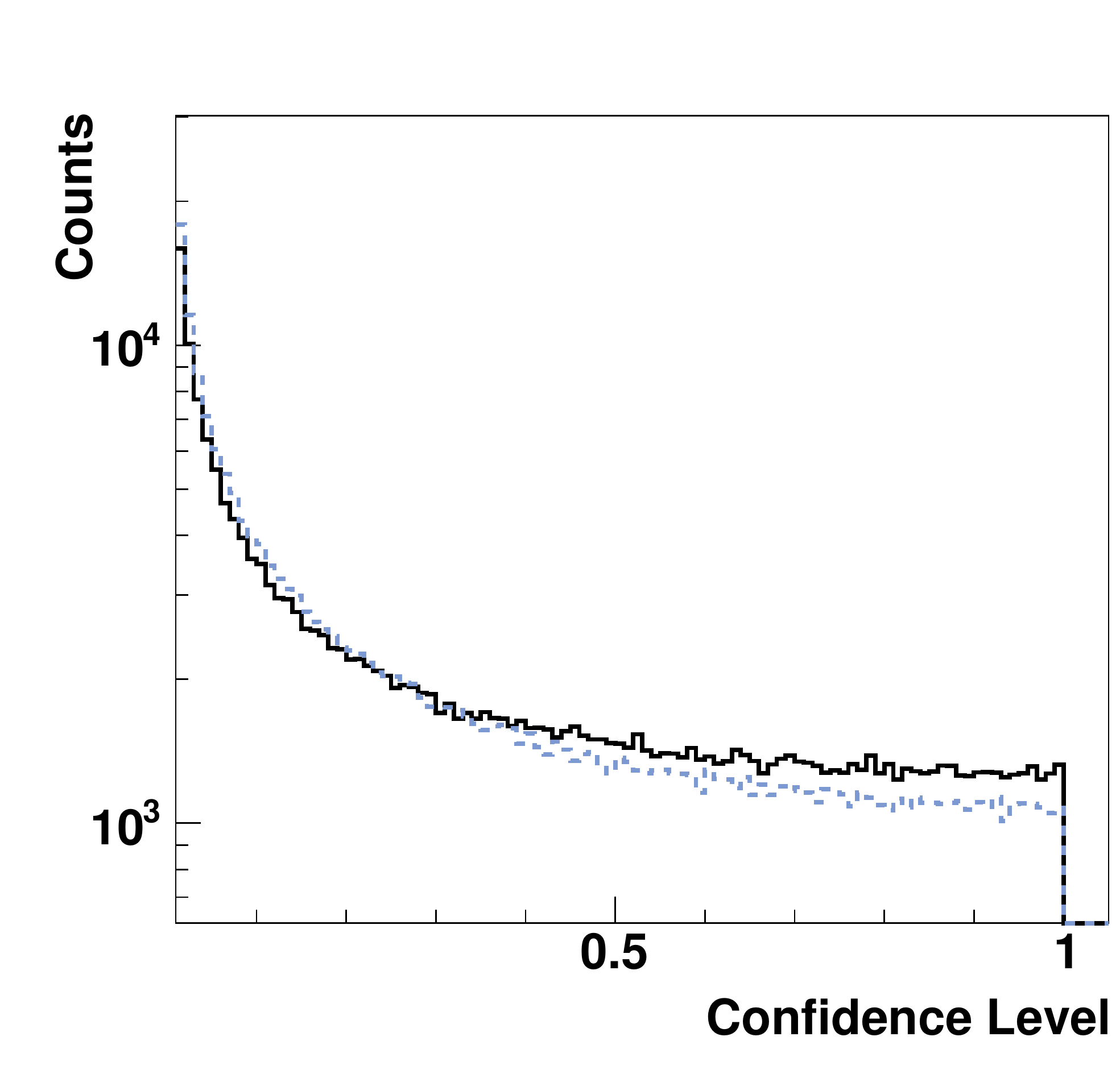}
 \caption{\label{Figure:cl}(Color online) Confidence-level distribution for a one-constraint (1C) fit testing 
   events for a missing $\pi^0$. The blue-dashed line is based on {\it raw} events, whereas the black-solid 
   line is based on the final event sample after all corrections.}
\end{figure}

In a second step, all events were subject to kinematic fitting. Events were tested for energy and momentum 
conservation in a four-constraint (4C) fit for detected particles and in a one-constraint (1C) fit for a missing
$\pi^0$. The exclusive reaction $\gamma p\to p\pi^+\pi^-$ was used to tune the covariance matrix in order 
to secure Gaussian pull distributions and a flat confidence-level (CL) distribution, where the confidence level 
denotes the goodness of fit to the data and is defined as the integral over the $\chi^2$~probability density 
function in the range $[\chi^2,\,\infty]$~\cite{brandt}. Figure~\ref{Figure:cl} shows confidence-level
distributions for the missing-$\pi^0$ hypothesis before (dashed-blue line) and after (solid-black line) all 
corrections. Events in this analysis were retained with a confidence-level cut of $p > 0.001$.

\begin{figure*}[t]
 \includegraphics[width=1.0\textwidth,height=0.24\textheight]{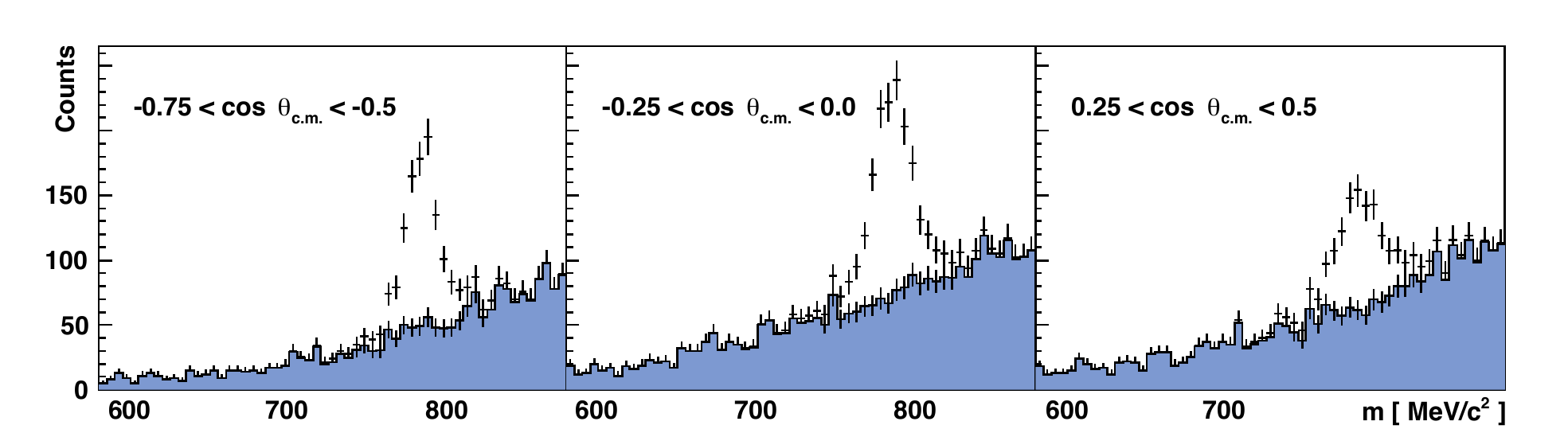}
 \caption{\label{Figure:MassDistributionsOmega}(Color online) Examples of invariant 
   $\pi^+\pi^-\pi^0$~mass distributions in the photon energy range $E_\gamma\in [1.5;\,1.6]$~GeV 
   for events that were subjected to the $Q$-factor fitting (background subtraction). These events 
   survived all kinematic cuts. 
   The solid blue area indicates the background.}
\end{figure*}

\subsection{\label{Subsection:Background}Background subtraction}
Frozen beads of butanol $\rm (C_{4}H_{9}OH)$ were used for the target material. When these butanol 
beads were polarized, only the 10~{\it free} hydrogen nucleons of the butanol could be polarized. Meson 
photoproduction on bound nucleons in $\rm ^{12}C$ and $\rm ^{16}O$ nuclei nonetheless generate a 
background beneath the signal from the polarized free nucleons. Owing to Fermi motion and final-state 
interactions, signals from reactions off $\rm ^{12}C$ and $\rm ^{16}O$ nucleons are broadened such that 
those signals do not form discernible peaks in the mass distributions. Although this background contribution 
drops out in the numerator of Eq.~(\ref{ReactionE}), the contribution from bound nucleons still remains in 
the denominator, requiring a procedure to remove any effects from that bound-nucleon contribution. Commonly, 
a dilution factor is calculated to account for the bound-nucleon contributions to the normalization in 
Eq.~(\ref{ReactionE}), defined as the ratio of the free-proton contribution to the full butanol cross section.
The energy- and angle-dependent effective dilution factors are usually determined from mass distributions
obtained from measurements on additional targets (such as the carbon and $\rm CH_{2}$ disk targets mentioned 
above in Sec.~\ref{Section:ExperimentalSetup}). However, in the measurements of the $\omega$~helicity 
asymmetry reported here, non-signal background events were removed in a probabilistic event-based approach
called the ``$Q$-factor method," described briefly here and detailed more fully in Ref.~\cite{Williams:2008sh}. 
That method was used for subtracting the background from the bound nucleons in the carbon and oxygen 
content of the butanol, as well as the removal of other sources of background.

The method assigns a quality factor (or $Q$ factor) to each event. These factors effectively serve as 
event-based dilution factors and describe the probability for an event to be a signal event. The $Q$~factors 
were then used to weight each event in the analysis when the observable was extracted. The method is a 
generalization of the traditional one-dimensional side-band subtraction method to higher dimensions without 
binning the data. Figure~\ref{Figure:MassDistributionsOmega} shows examples of the resulting separation of 
signal and background in the invariant $\pi^+\pi^-\pi^0$ mass distribution. Three angle bins are presented 
in the energy range $E_\gamma \in [\,1.5; 1.6\,]$~GeV. The sum of the signal (white area) and the background
(blue area) is identical to the total unweighted mass distribution, whereas the invariant $3\pi$~mass of 
each event weighted by $1-Q$ gives the background alone.

In this event-based method, the general set of coordinates that describe the multi-dimensional phasespace 
of a reaction is separated into {\it reference} and {\it non-reference} coordinates. In this analysis, the 
invariant $M_{\pi^+\pi^-\pi^0}$ mass was chosen as the reference coordinate. The $Q$-factor method proceeded 
with the selection of the $N_c$ kinematically-nearest neighbors for each event. A number of $N_c=300$ was 
chosen by defining a distance metric for the individual kinematic variables spanning the phase~space:
\begin{equation}
  D_{ab}^2 \,=\,\sum_{i=1}^5 \bigg( \frac{\Gamma^a_i\,-\,\Gamma^b_i}{\Delta_i}\bigg)^2\,,
  \label{equidistance}
\end{equation}
where the $\Gamma_i$ denote the set of kinematic variables for the two events $a$ and $b$, and $\Delta_i$ 
is the full range for the kinematic variable $i$. The following independent non-reference variables were used:
\begin{equation}
{\rm cos}\,\theta^{\,\omega}_{\rm c.m.},~{\rm cos}\,\theta_{\rm \,HEL},~\phi_{\rm \,HEL},~\phi^{\,\omega}_{\rm lab},~\lambda\,,
\end{equation}
where cos$\,\theta^{\,\omega}_{\rm c.m.}$ denotes the cosine of the polar angle of the $\omega$ in the 
center-of-mass frame, cos$\,\theta_{\rm HEL}$ and $\phi_{\rm HEL}$ are the two angles of the $\omega$
in the helicity frame, and $\phi^{\,\omega}_{\rm lab }$ is the azimuthal angle of the $\omega$ in the laboratory 
frame. Defined in terms of the pion momenta in the $\omega$ rest frame, the variable 
$\lambda = |\,\vec{p}_{\pi^+}\,\times\,\vec{p}_{\pi^-}|^2 \,/\,\lambda_{\rm \,max}$ is proportional to the 
$\omega\to\pi^+\pi^-\pi^0$ decay amplitude as a consequence of isospin conservation~\cite{Williams:2009ab}, 
with $\lambda_{\rm \,max}$ defined as~\cite{Weidenauer:1993mv}
\begin{equation}
\lambda_{\rm \,max} \,=\, Q^2\,\bigg(\frac{Q^2}{108}\,+\,\frac{mQ}{9}\,+\,\frac{m^2}{3} \bigg)
\end{equation}
for a totally symmetric decay, where $Q = T_1 + T_2 + T_3$ is the sum of the $\pi^{\pm,\,0}$~kinetic energies 
and $m$ is the $\pi$~mass. The parameter~$\lambda$ varies between 0 and 1 and shows a linearly-increasing 
distribution as expected for a vector meson.
Event-based maximum likelihood fits were performed of the invariant $M_{3\pi}$~distributions for
every selected event and its $N_{c}$ nearest neighbor events according to:
\begin{equation}
  f(x) \,=\, N\,\cdot [ f_{s}\,\cdot\, S(x) \,+\, ( 1 \,-\, f_{s} )\,\cdot\, B(x) ]\,,
  \label{equ:total_function}
\end{equation}
where $S(x)$ and $B(x)$ denote the signal and the background probability density functions, respectively, 
and $x = M_{3\pi}$. A Voigt profile was chosen for the signal and the background shape was modeled with a 
second-order Chebychev polynomial. The para\-meter~$N$ was a normalization constant and $f_{s}$ was the 
signal fraction with a value between 0 and 1. The $Q$~factor is defined by:
\begin{equation}
  Q \,=\, \frac{s(x)}{s(x) \,+\, b(x)}\,,
  \label{equ:Q_factor}
\end{equation}
where $x$ is again the invariant mass of the $\pi^+\pi^-\pi^0$~system, $s(x) = f_{s} \cdot S(x)$, and 
$b(x) = (1-f_{s}) \cdot B(x)$.

\section{\label{Section:Observable}Extraction of the $E$~Observable}
Data using an unpolarized- or a circularly-polarized photon beam in combination with an unpolarized- 
or a longitudinally-polarized target are isotropic in the laboratory azimuthal angle since the orientation 
of any particle polarization is along the $z$-axis in the laboratory frame. Any polarization asymmetry 
for a kinematic bin is given by the difference in the event counts for parallel/anti-parallel polarization 
settings:
\begin{equation}
A^\Rightarrow\,=\,\frac{N^\Rightarrow_\leftarrow\,-\,N^\Rightarrow_\rightarrow}
                                  {N^\Rightarrow_\leftarrow\,+\,N^\Rightarrow_\rightarrow}~=~
A^\Leftarrow\,=\,\frac{N^\Leftarrow_\rightarrow\,-\,N^\Leftarrow_\leftarrow}
                                 {N^\Leftarrow_\rightarrow\,+\,N^\Leftarrow_\leftarrow}~,
\label{eq:asymmetry}
\end{equation}
where $\rightarrow$~($\leftarrow$) and $\Rightarrow$~($\Leftarrow$) indicate if the photon 
and nucleon spin points downstream (upstream), respectively.

The corresponding polarization observable can then be extracted from this asymmetry and 
Eq.~(\ref{ReactionE}) reduces to:
\begin{equation}
E\,=\,\frac{1}{\Lambda_z^\Rightarrow\,\delta_\odot}\,A^\Rightarrow\,=\,
          \frac{1}{\Lambda_z^\Leftarrow\,\delta_\odot}\,A^\Leftarrow~\,,
\label{eq:observable}
\end{equation}
where $\delta_\odot$ denotes the degree of circular photon-beam polarization and $\Lambda_z$ is 
the degree of longitudinal target-proton polarization

\section{\label{Section:Results}Experimental Results}
The kinematics of $\omega$~photoproduction from the proton can be completely described by two kinematic 
variables. The incoming photon energy $E_\gamma$ and cos\,$\theta^{\,\omega}_{\rm c.m.}$ were chosen, 
where $\theta^{\,\omega}_{\rm c.m.}$ is the polar angle of the photoproduced $\omega$ in the 
center-of-mass frame. The $z$-axis was defined as the direction of the incident photon beam.

\subsection{The $E$ observable for $\gamma p\to p\omega$}
\begin{figure*}[t]
 \includegraphics[width=0.98\textwidth,height=0.55\textheight]{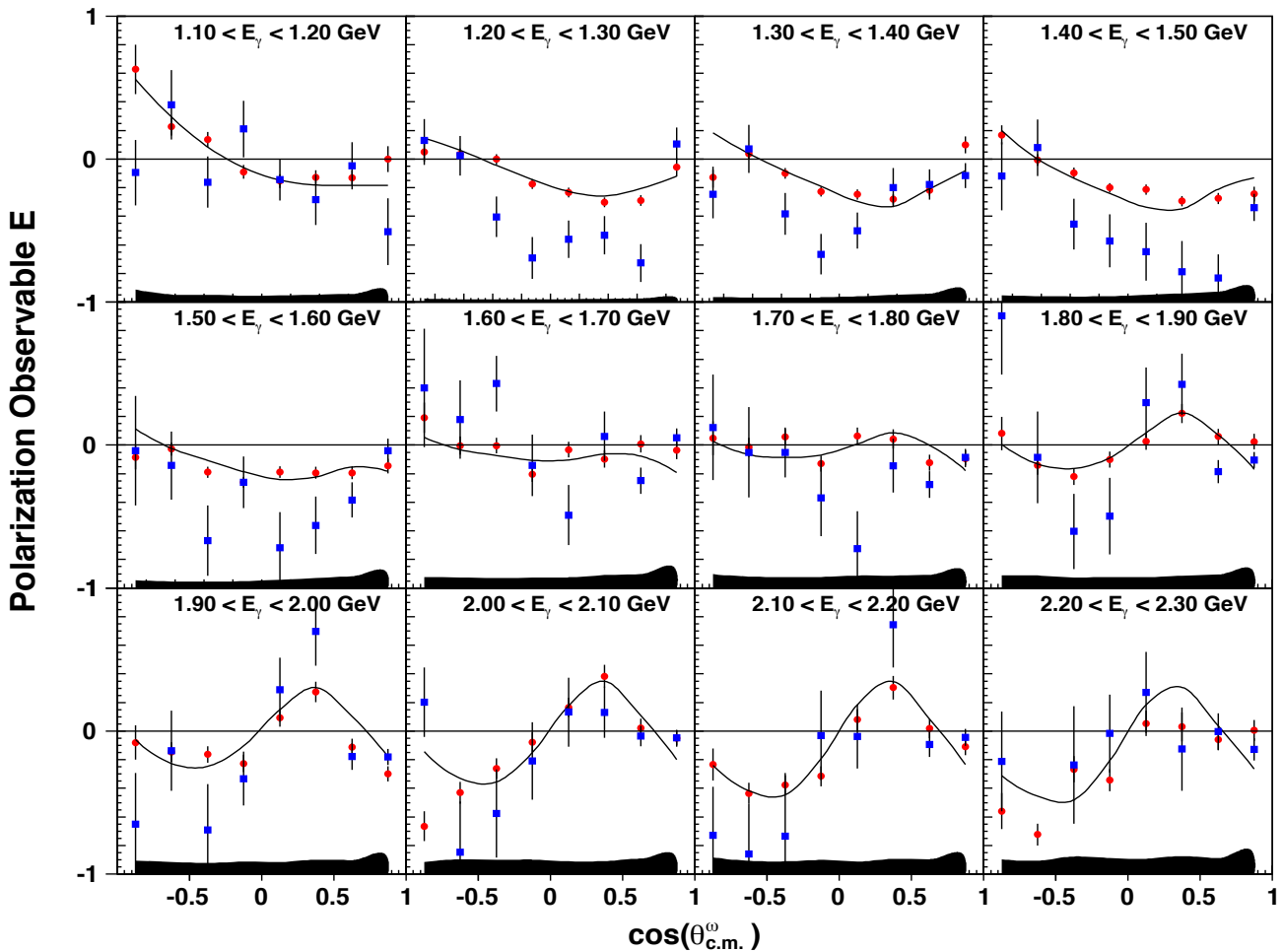}
  \vspace{-0.1cm}
  \caption{\label{Figure:OmegaResultsEnergyBins}(Color online) Measurement of the helicity asymmetry~$E$
    in the reaction $\gamma p\to p\omega$ using a circularly-polarized photon beam and 
    a longitudinally-polarized target. The data are shown in 100-MeV-wide bins for the photon 
    energy range $E_\gamma\in [1.1,\,2.3]$~GeV. The CLAS-FROST results (red circles
    {\color{red} $\bullet$}) are compared with results from the CBELSA/TAPS 
    Collaboration~\cite{Eberhardt:2015lwa}, which used the radiative decay mode, 
    $\omega\to \pi^0\gamma$~(blue squares {\tiny\color{blue} $\blacksquare$}).
    The black solid line represents the BnGa PWA solution. The data points include statistical 
    uncertainties only; the total systematic uncertainty is given as bands at the bottom of each distribution.}
\end{figure*}

\begin{figure*}[t]
 \includegraphics[width=1.02\textwidth]{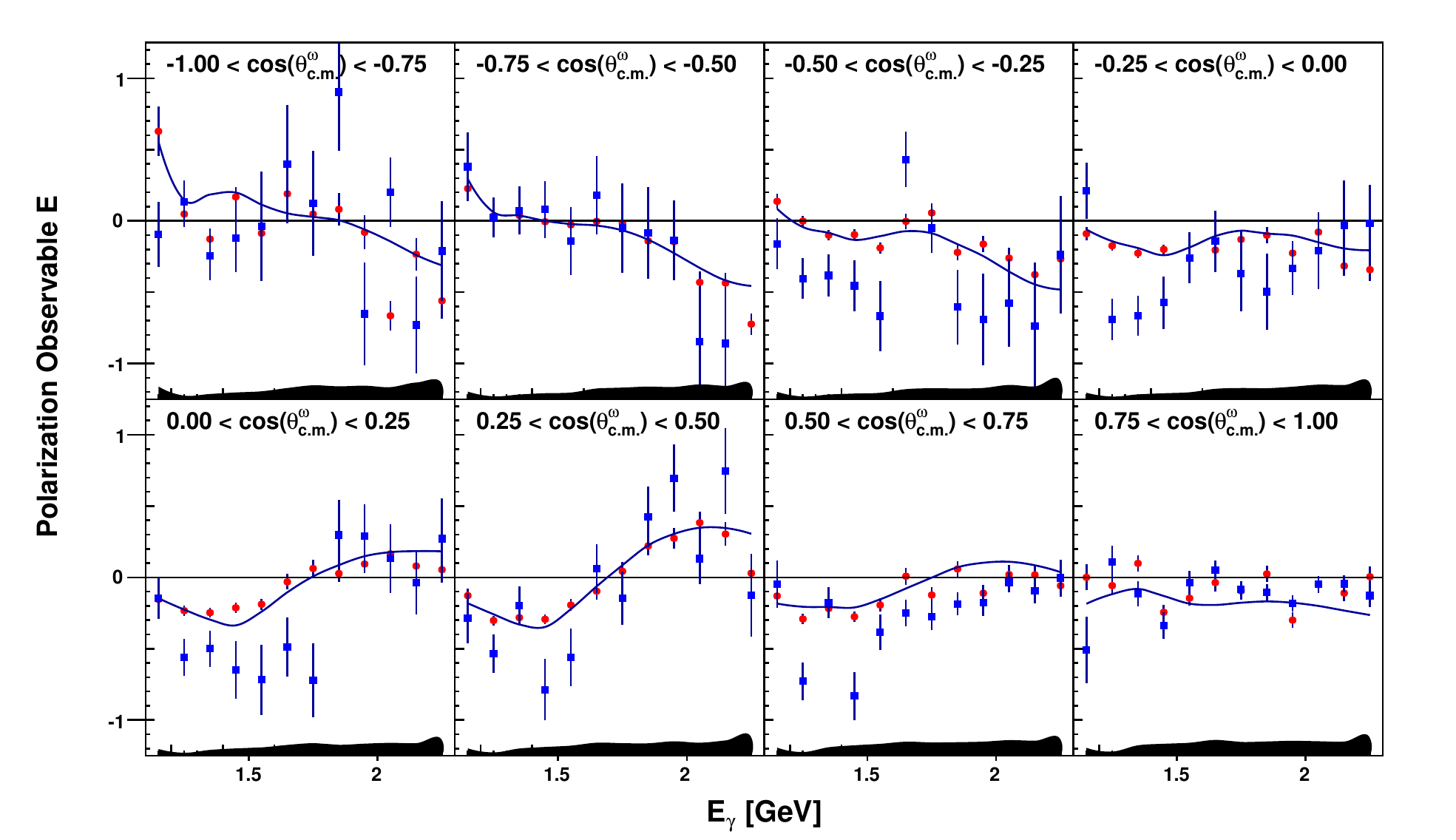}
 \caption{\label{Figure:OmegaResultsAngleBins}(Color online) Measurement of the helicity asymmetry~$E$
    in the reaction $\gamma p\to p\omega$ using a circularly-polarized photon beam and a
    longitudinally-polarized target. The data are shown in 0.25-wide-bins in cos\,$\theta^{\,\omega}_{\rm c.m.}$
    and in 100-MeV-wide bins for the photon energy range $E_\gamma\in [1.1,\,2.3]$~GeV. The CLAS-FROST 
    results (red circles {\color{red} $\bullet$}) are compared with results from the CBELSA/TAPS 
    Collaboration~\cite{Eberhardt:2015lwa}, which used the radiative decay mode, 
    $\omega\to \pi^0\gamma$~(blue squares {\tiny\color{blue} $\blacksquare$}).
    The black solid line represents the BnGa PWA solution. The data points include statistical 
    uncertainties only; the total systematic uncertainty is given as bands at the bottom of each distribution.}
\end{figure*}

Figure~\ref{Figure:OmegaResultsEnergyBins} shows the $E$~observable for the photoproduction of a 
single-$\omega$ meson off the proton from this analysis (red circles {\color{red} $\bullet$}). The 
angular distributions are shown for 100-MeV-wide bins in the incoming photon energy. 
Figure~\ref{Figure:OmegaResultsAngleBins} shows the energy dependence of the $E$~observable 
for eight angle bins in cos\,$\theta^{\,\omega}_{\rm c.m.}$. For comparison, earlier results from the 
CBELSA/TAPS Collaboration~\cite{Eberhardt:2015lwa} are also shown (blue boxes {\tiny\color{blue} 
$\blacksquare$}) in Figs.~\ref{Figure:OmegaResultsEnergyBins} and~\ref{Figure:OmegaResultsAngleBins}. 
Both data sets are consistent in their asymmetry behavior (same sign for almost every data point). However,
larger discrepancies in the magnitude are visible, in particular at low energies, $E_\gamma < 1.5$~GeV. 

In an effort to resolve these discrepancies, we identified three likely sources: (1) beam polarization, (2) 
target polarization, and (3) background subtraction (see also Eqs.~(\ref{eq:asymmetry}) and (\ref{eq:observable})). 
The values for the accelerator-beam polarization and the target polarization used in this analysis are the 
same as those values applied in the extraction of the $\eta$~helicity asymmetry at CLAS described in 
Ref.~\cite{Senderovich:2015lek}. This CLAS-$\eta$ analysis was based on the same FROST data set as the
$\omega$~analysis presented here. The $\eta$~observable showed the expected flat behavior close to the 
reaction threshold and a magnitude of almost one owing to the dominance of the $N(1535)\,1/2^-$ nucleon 
resonance. Moreover, the mass distributions presented in Fig.~\ref{Figure:MassDistributionsOmega}, which 
refer to the incident-photon energy range $[1.5,\,1.6]$~GeV (see Fig.~\ref{Figure:OmegaResultsEnergyBins}), 
do not indicate that our background-subtraction technique is the major cause for the observed discrepancy 
between the CLAS and the CBELSA/TAPS results in this energy range. We note that a possible overestimation 
of the $\omega\to\pi^0\gamma$ yields at ELSA may be the origin of the inconsistency between the two data 
sets. For the radiative decay of the $\omega$, the reactions $\gamma p\to p\pi^0\pi^0$ (with one low-energy 
photon undetected) and $\gamma p\to p\pi^0$ (with an additional photon misrepresented), exhibit ``peaking'' 
background close to the $\omega$ in the invariant $\pi^0\gamma$ mass distribution, which is very challenging 
to account for. We refer to Refs.~\cite{Wilson:2015uoa,Eberhardt:2015lwa} for more details on the techniques of 
analyzing the reaction $\gamma p\to p\omega\to p\pi^0\gamma$.

\subsection{\label{Subsection:SytematicUncertainties}Systematic uncertainties}
The statistical uncertainties were determined from the number of events in each ($E_\gamma$, 
cos\,$\theta^{\,\omega}_{\rm c.m.}$) bin and are shown for all data points in Figs.~\ref{Figure:OmegaResultsEnergyBins} 
and~\ref{Figure:OmegaResultsAngleBins}; systematic uncertainties are given as bands at 
the bottom of each distribution.

The overall systematic uncertainty includes uncertainties in the degree of photon-beam and 
target-proton polarization, contributions from the electron-beam charge asymmetry, and the 
background-subtraction method. The systematic uncertainties in the degree of photon-beam 
and target-proton polarization are applied as global factors of 2\,\% and 3\,\%, respectively. 
Other sources of systematic uncertainty are described below.

The electron-beam polarization was toggled between the $h^+$ and the $h^-$~helicity states at a rate
of 29.560~Hz. At these large rates, the photon-beam flux for both helicity states should be the same,
on average. However, small beam-charge asymmetries of the electron beam can cause instrumental 
asymmetries and lead to systematic deviations in the hadronic asymmetries. The electron beam-charge 
asymmetry was calculated from the luminosities of $h^+$ and $h^-$~events:
\begin{equation}
\Gamma^\pm\,=\,\alpha^\pm\,\Gamma\,=\,\frac{1}{2}\,(1\,\pm\,\bar{a}_c)\,\Gamma\,,
\end{equation}
where $\Gamma$ was the total luminosity and $\alpha^\pm$ denoted the fraction of $h^+$ and 
$h^-$~events. The parameter $\alpha^\pm$ depended on the mean value of the electron beam-charge 
asymmetry, $\bar{a}_c$, which was typically less than 0.2\,\%. Therefore, contributions from this source 
of the systematic uncertainty were considered negligible. 

The $Q$-factor method will lead to a certain level of correlation among events because events can 
share a significant number of the same neighbors in the limit of very small statistics. For this reason, 
the systematic uncertainty in the $\omega$~yield in a kinematic bin due to the $Q$-factor method was 
obtained from the covariance matrix of each fit and the correlation factors between events $i$ and $j$, 
which describe the fraction of shared nearest neighbor events between two events. The systematic 
variance is given by 
\begin{equation}
\sigma^2_{\omega}\,=\sum_{i,j}\, \sigma_Q^i\,\rho_{ij}\,\sigma_Q^j\,,
\end{equation}
where the sum $i, j$ extends over all events in a kinematic ($E_\gamma$, cos\,$\theta^{\,\omega}_{\rm c.m.}$)~bin, 
$\sigma_Q^i$ and $ \sigma_Q^j$ are the fit uncertainties for events $i$ and $j$, and $\rho_{ij}$ is the correlation
factor between events $i$ and $j$. The absolute uncertainties due to the $Q$-factor method 
range from about 0.03 close to the reaction threshold to about 0.1 at $E_\gamma = 2$~GeV.

An additional possible source of systematic uncertainty is the presence of accidental photons. The 
fraction of accidental photons was at most 2.5\,\%. It was estimated from comparing the central 
peak with the neighboring electron beam buckets in the {\it coincidence-time} spectrum, which is
defined per photon as the difference between the Tagger time and the Start Counter time at the 
interaction point. Accidental photons lead to an overestimation of the event numbers but drop 
out in the asymmetry of event counts.

\section{\label{Section:PWA}Partial-Wave Analysis}
Baryon resonances are very short-lived and as a result, these states are broad and overlapping. 
Contributions from resonances manifest themselves as enhancements or peak-like structures 
in the cross sections. Owing to the broad nature of baryon resonances however, peaks in the 
experimental cross sections are typically based on contributions from many resonances and 
can merely be addressed as resonance regions. Extracting $N^\ast$~parameters from the data 
thus remains a challenge. Amplitude analyses or PWAs need to be performed in order to identify
resonance contributions in a particular reaction. The situation becomes more complicated at higher 
resonance masses because many reaction channels need to be considered. Any reliable extraction of 
resonance properties must therefore be based on a coupled-channels approach. 

In recent years, several groups have contributed significantly to our understanding of the baryon spectrum, 
but a comprehensive analysis based on a large database of observables has been performed only at a
few institutions, see e.g. Refs.~\cite{Klempt:2009pi,Crede:2013sze} and references therein. The precise 
photoproduction data resulting from recent experiments have a great significance for the extraction of 
baryon resonance parameters. In particular, the data on some polarization observables are decisive in 
avoiding ambiguities in the description of resonance contributions. 

This section describes the results of a PWA in the framework of the BnGa coupled-channels approach 
that is based on the new $\gamma p\to p\omega$~data from CLAS on the polarization observables~$E$ 
(presented here), $T$ and $\Sigma$~\cite{Roy:2017qwv,Collins:2017vev}, and $F$, $P$, and $H$~\cite{FROST:PRL}. 
The CLAS data were added to the full BnGa database, which includes a large set of data on pion- 
and photo-induced meson-production reactions, with up to two pseudoscalar mesons in the final 
state~\cite{BnGa:Database}. The BnGa group has recently reported on a PWA~\cite{Denisenko:2016ugz} 
that, at the time, was restricted to $\omega$~data from the CBELSA/TAPS experiment alone: 
(1) Differential cross sections and spin-density matrix elements (SDMEs) for unpolarized incident 
photons ($\rho^0_{00}$, $\rho^0_{10}$, $\rho^0_{1\,-1}$) covering the energy range from threshold 
to 2500~MeV, as well as SDMEs for linearly-polarized incident photons ($\rho^1_{00}$, $\rho^1_{11}$, 
$\rho^1_{1\,-1}$, $\rho^1_{10}$, $\rho^2_{10}$, $\rho^2_{1\,-1}$) covering the energy range 
$E_\gamma < 1650$~MeV~\cite{Wilson:2015uoa}; (2) Data on the photon-beam asymmetry 
$\Sigma$~\cite{Klein:2008aa}; (3) Results on the helicity asymmetry, $E$, ($E_\gamma < 2300$~MeV) 
and the $G$~observable for one bin in photon energy ($1108 < E_\gamma < 1300$~MeV)~\cite{Eberhardt:2015lwa}.

The new BnGa-PWA solution, which is based on the CLAS data, is shown in 
Fig.~\ref{Figure:OmegaResultsEnergyBins} and~\ref{Figure:OmegaResultsAngleBins} 
as a solid line. More details on the PWA and branching ratios for $N^\ast$ decays into $N\omega$ 
will be discussed in a subsequent publication~\cite{Anisovich:2017}. The inclusion of SDMEs allowed 
the study of the $\omega$~production process in more detail and helped separate the natural and 
unnatural parity-exchange contributions. In the BnGa analysis, $\pi$~exchange in the $t$-channel 
was found to remain small across the analyzed energy range, while pomeron $t$-channel exchange 
gradually grew from the reaction threshold to dominate all other contributions above $W \approx 2$~GeV
($E_\gamma > 1.66$~GeV).


In the BnGa analysis, close to the reaction threshold, $J^P = 3/2^+$ remains the leading resonant 
partial wave and shows a strong peak with a maximum around $W = 1.8$~GeV. This wave is identified 
with the $N(1720)\,3/2^+$~state, which is situated just below the reaction threshold. The $3/2^+$
partial wave has a more complex structure and indications for at least one more resonance around 
$W = 1.9$~GeV have been found. The contributions from the $1/2^-$ and $3/2^-$ partial waves are 
notably smaller compared to the leading $3/2^+$ partial wave. The $J^P = 1/2^-$~wave has a 
maximum close to the reaction threshold, which can be identified with the $N(1895)\,1/2^-$~resonance, 
and smoothly declines toward higher masses; no further structures are observed. The $J^P = 3/2^-$~wave 
reaches a maximum just above 2~GeV, which can be identified with contributions from the $N(2100)\,3/2^-$~state. 
The $J^P = 5/2^+$~wave exhibits a richer structure. This wave has a local enhancement close to the 
threshold, identified with $N(1680)\,5/2^+$, and a maximum around $W = 2$~GeV; the latter is identified 
with the poorly-established $N(2000)\,5/2^+$ state. The $N\omega$~coupling of this resonance has 
significantly increased compared to the previous BnGa $\omega$~PWA. The contributions from the 
$5/2^-$, $7/2^+$, and $7/2^-$ partial waves remain smaller. In all fits, they were found to be less 
than about~5\,\%. The 7/2 partial waves play an important role in the description of the density 
matrices at masses above~2.1~GeV.

\section{\label{Section:Summary}Summary}
The double-polarization observable~$E$ for the reaction $\gamma p\to p\,\omega$ has been 
measured at CLAS using the frozen-spin FROST target, covering the energy range from 1.1~to 
2.3~GeV using the $\omega\to\pi^+\pi^-\pi^0$ decay. Fairly large helicity asymmetries are 
observed, indicating significant contributions from $s$-channel $N^{\ast}$~resonances. The data 
have been partial-wave analyzed within the BnGa coupled-channels framework and contributions 
from $N^\ast$~resonances have been identified. The leading partial waves at the reaction threshold 
are the $3/2^+$ and $5/2^+$~waves. Toward higher energies around $W\approx 2$~GeV, the $t$-channel 
contributions increase in strength and are defined by a dominant pomeron exchange and a smaller $\pi$ 
exchange. In addition, further contributions from nucleon resonances are required to describe the data. 
The $1/2^-$, $3/2^-$, and $5/2^+$ partial waves show considerable contributions to the PWA solution.

\begin{acknowledgments}
The authors thank the technical staff at Jefferson Lab and at all the participating 
institutions for their invaluable contributions to the success of the experiment. 
This material is based upon work supported by the U.S. Department of Energy, 
Office of Science, Office of Nuclear Physics, under Contract No. DE-AC05-06OR23177.
This work was also supported by the US National Science Foundation, the State 
Committee of Science of Republic of Armenia, the Chilean Comisi\'{o}n Nacional de 
Investigaci\'{o}n Cient\'{i}fica y Tecnol\'{o}gica (CONICYT),
the Italian Istituto Nazionale di Fisica Nucleare, the French Centre National de la 
Recherche Scientifique, the French Commissariat a l'Energie Atomique, the Scottish 
Universities Physics Alliance (SUPA), the United Kingdom's Science and Technology 
Facilities Council, the National Research Foundation of Korea, the Deutsche 
Forschungsgemeinschaft (SFB/TR110), and the Russian Science Foundation
under Grant No. 16-12-10267. 
\end{acknowledgments}


\end{document}